\documentclass[aps,pre,floats,floatfix,twocolumn,showpacs,superscriptaddress,nofootinbib]{revtex4}

\usepackage{latexsym,amsmath,amssymb}
\usepackage{amsbsy}
\usepackage[pdftex]{graphicx}
\usepackage{epstopdf}
\usepackage[tight]{subfigure}

\renewcommand{\geq}{\geqslant}

\begin{document}

\title{Curvature and Temperature of Complex Networks}

\author{Dmitri Krioukov}
\author{Fragkiskos Papadopoulos}
\affiliation{Cooperative Association for Internet Data Analysis (CAIDA),
University of California-San Diego (UCSD), La Jolla, California 92093, USA}
\author{Amin Vahdat}
\affiliation{Department of Computer Science and Engineering,
University of California-San Diego (UCSD), La Jolla, California 92093, USA}
\author{Mari{\'a}n Bogu{\~n}{\'a}}
\affiliation{Departament de F{\'\i}sica Fonamental, Universitat de
  Barcelona, Mart\'{\i} i Franqu\`es 1, 08028 Barcelona, Spain}

\begin{abstract}

We show that heterogeneous degree distributions in
observed scale-free topologies of complex networks can emerge as a
consequence of the exponential expansion of hidden hyperbolic
space. Fermi-Dirac statistics provides a physical interpretation of
hyperbolic distances as energies of links. The hidden space
curvature affects the heterogeneity of the degree distribution,
while clustering is a function of temperature. We embed the Internet
into the hyperbolic plane, and find a remarkable congruency between
the embedding and our hyperbolic model. Besides proving our model
realistic, this embedding may be used for routing with only local
information, which holds significant promise for
improving the performance of Internet routing.

\end{abstract}


\maketitle

Many complex networks possess heterogeneous degree distributions.
This heterogeneity is often modeled by power laws, often
truncated~\cite{ClaSha09}. These networks also exhibit strong
clustering, i.e., high concentration of triangular subgraphs. Our
previous work~\cite{SeKrBo08} demonstrated that the clustering
peculiarities of complex networks, and in particular their
self-similarity, finds a natural geometric explanation in the
existence of hidden metric spaces underlying the network and
abstracting the intrinsic similarities between its nodes. Here we
seek to provide a geometric interpretation of the first
property---network heterogeneity. We show that heterogeneous, or
scale-free, degree distributions in complex networks appear as a
simple consequence of negative curvature of hidden spaces. That is,
we argue that these spaces are hyperbolic.

\begin{figure}[!]
    \subfigure{
    \includegraphics[width=1.63in]{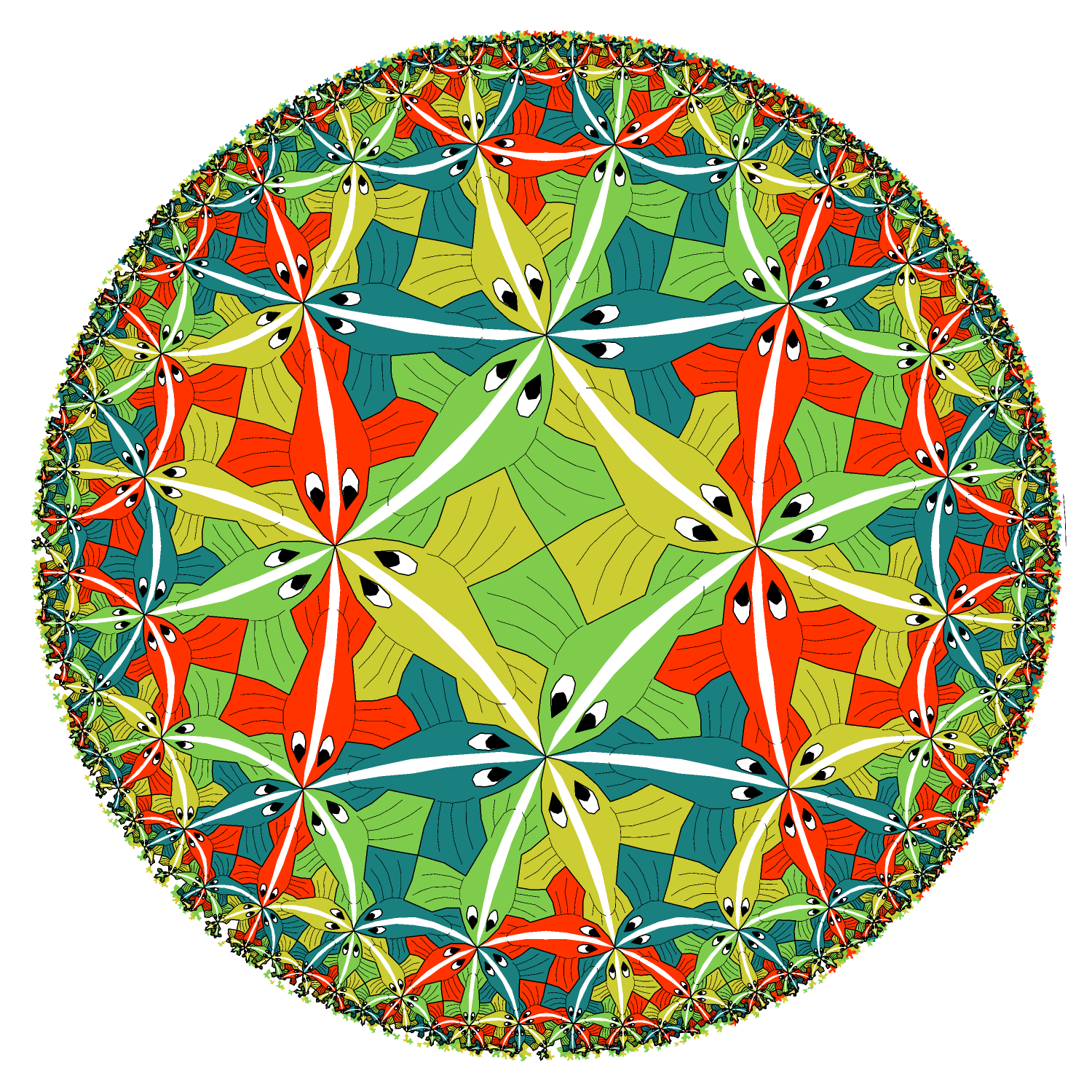}
    }\hfill
    \subfigure{
    \includegraphics[width=1.63in]{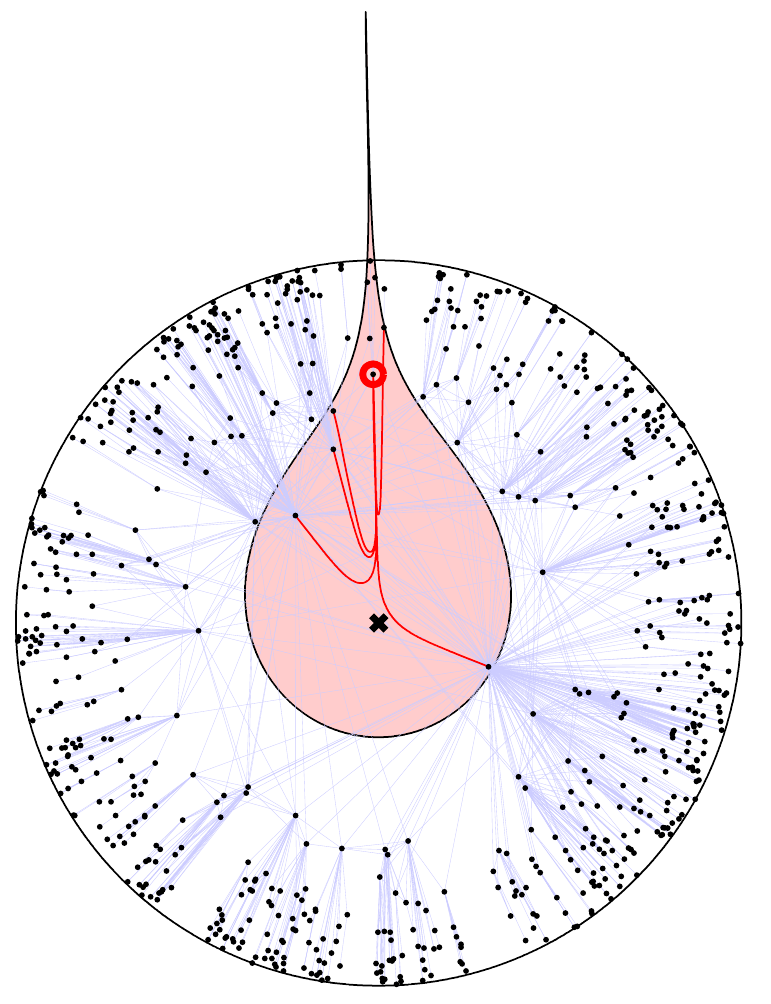}}
    \caption{
    {\bf Left:}
    Artistic visualization of the Poincar\'e disc model of
    the hyperbolic plane $\mathbb{H}^2$ by Levy, based on Escher's
    {\it Circle Limit III}, with
    the permission from the Geometry Center, University
    of Minnesota.
    The exponential expansion of fish illustrates the exponential
    expansion of hyperbolic space. All fish are of the same hyperbolic
    size, but their Euclidean size exponentially decreases, while
    their number exponentially increases with the distance from the
    origin.
    {\bf Right:}
    A modeled network with $N=740$ nodes,
    power-law exponent $\gamma=2.2$,
    and average degree $\bar{k}\approx 5$
    embedded in the hyperbolic disc of curvature $K=-1$ and radius $R\approx15.5$.
    The Euclidean distance between a node and the origin at the disc center, shown as the
    cross, represents the true hyperbolic distance between the two.
    But the Euclidean distance between any two other nodes is \emph{not} equal
    to the hyperbolic distance between them, as indicated by the peculiar shape
    of the shaded hyperbolic \emph{disc} centered at the circled node located
    at distance $r=10.6$ from the origin. The hyperbolic radius of this disc is also $R$,
    and according to the model, the circled node is connected to all the nodes lying in this disc.
    The curves show the hyperbolically straight lines, i.e., geodesics, connecting
    the circled node and some nodes in its disc.
    \label{fig:escher}
    }
\end{figure}

The main metric property of hyperbolic geometry is the exponential
expansion of space, see Fig.~\ref{fig:escher}, left. For example, in
the hyperbolic plane, i.e., the two-dimensional space of constant
curvature $-1$, the length of a circle and the area of a disc of
radius $R$ are $2\pi\sinh R$ and $2\pi(\cosh R -1)$, both growing as
$\sim e^R$. The hyperbolic plane is thus metrically equivalent to an
$e$-ary tree, i.e., a tree with the average branching factor equal
to $e$. Indeed, in a $b$-ary tree the surface of a sphere or the
volume of a ball of radius $R$, measured as the number of nodes
lying at or within $R$ hops from the root, grow as $b^R$.
Informally, hyperbolic spaces can therefore be thought of as
``continuous versions'' of trees.

To see why this exponential expansion of hidden space is intrinsic
to complex networks, observe that their topology represents the
structure of connections or interactions among distinguishable,
heterogeneous elements abstracted as nodes. This heterogeneity
implies that nodes can be somehow classified, however broadly, into
a taxonomy, i.e., nodes can be split into large groups consisting of
smaller subgroups, which in turn consist of even smaller
subsubgroups. The relationships between such groups and subgroups
can be approximated by tree-like structures, sometimes called
\emph{dendrograms}, in which the distance between two nodes
estimates how similar they are~\cite{WatDoNew02}. Importantly, the
node classification hierarchy need not be strictly a tree.
Approximate ``tree-ness,'' which can be formally expressed solely in
terms of the metric structure of a space~\cite{Gromov07-book}, makes
the space hyperbolic.

Let us see what network topologies emerge in the simplest possible
settings involving hidden hyperbolic metric spaces. Let us form a
network of $N\gg1$ nodes located in the hyperbolic plane
$\mathbb{H}^2$. Since the number of nodes is finite, the area that
nodes occupy is bounded. Let $R\gg1$ be the radius of a disc within
which nodes are uniformly distributed. In hyperbolic geometry, this
means that nodes are given an angular coordinate $\theta$ randomly
distributed in $[0,2\pi]$, and a radial coordinate $r$ following the
density $\rho(r) = \sinh r/(\cosh R-1) \approx e^{r-R}$. Next, we
have to specify the connection probability $p(x)$ that two nodes at
hyperbolic distance $x$ are connected. We first consider the
simplest case, the step function $p(x) = \Theta(R-x)$, and justify
this choice later. This $p(x)$ connects each pair of nodes if the
hyperbolic distance between them is not larger than $R$.

The network is now formed, and we can compute the average degree
$\bar{k}(r)$ of nodes at distance $r$ from the disc center. These
nodes are connected to all nodes in the intersection area of the two
discs of the same radius $R$, one in which all nodes reside, and the
other centered at distance $r$ from the center of the first disc,
see Fig.~\ref{fig:escher}, right. Since the node distribution is
uniform, $\bar{k}(r)$ is proportional to the area of this
intersection, which decreases exponentially with $r$, $\bar{k}(r)
\sim e^{-r/2}$. Therefore, the inverse function is logarithmic,
$\bar{r}(k) \sim -2 \ln k$, and the node degree distribution in the
network is approximately a power law, $P(k) \approx
\rho[\bar{r}(k)]\,|\bar{r}'(k)| \sim k^{-3}$. If we generalize the
space curvature to $K=-\zeta^2$, $\zeta>0$, and the node density to
$\rho(r) \approx \alpha e^{\alpha(r-R)}$, where we can think of
$\alpha>0$ as the logarithm of the average branching factor in the
underlying hierarchy, then the average degree at radius $r$ scales
as $\bar{k}(r) \sim e^{-\zeta r/2}$ if $\alpha/\zeta\geq1/2$, or
$\bar{k}(r) \sim e^{-\alpha r}$ otherwise, so that the node degree
distribution becomes $P(k) \sim k^{-\gamma}$ with
\begin{equation}\label{eq:true-gamma}
\gamma =
\begin{cases}
2\alpha/\zeta+1 & \text{if $\alpha/\zeta\geq1/2$,}\\
2 & \text{otherwise}.
\end{cases}
\end{equation}
To fix the average degree in the network, we have to choose
$N=c\,e^{\frac{\zeta}{2}R}$, where $c$ is a constant. The result in
Eq.~(\ref{eq:true-gamma}) is remarkable as it shows that
heterogeneous degree distributions may emerge as a simple consequence
of the exponential expansion of hyperbolic space.

However, our choice of the step-function connection probability is
not yet justified. To justify it, and to show that scale-free
networks have effective hyperbolic geometries underneath, we recall
the $\mathbb{S}^1$ model introduced in~\cite{SeKrBo08}. In that
model, networks are constructed as follows. First, distribute $N$
nodes uniformly over the circle $\mathbb{S}^1$ of radius $N/(2\pi)$,
so that the node density on the circle is fixed to $1$. Second,
assign to all nodes an additional hidden variable $\kappa$
representing their expected degrees. To generate scale-free
networks, the variable $\kappa$ is power-law distributed according
to $\rho(\kappa)=\kappa_0^{\gamma-1}(\gamma-1)\kappa^{-\gamma}$,
$\kappa \in [\kappa_{0},\infty)$, where $\kappa_0$ is the minimum
expected degree. Finally, let $\kappa$ and $\kappa'$ be the expected
degrees of two nodes located at distance $d=N\Delta\theta/(2\pi)$
measured over the circle ($\Delta\theta$ is the angular distance
between the nodes). We connect each pair of nodes with probability
$\widetilde{p}(\chi)$, where $\chi \equiv d/(\mu \kappa \kappa')$,
and constant $\mu$ fixes the average degree in the network.

The key point is that the connection probability
$\widetilde{p}(\chi)$ can be \emph{any integrable function}. As long
as the distance over the circle is rescaled as $\chi \sim d/(\kappa\kappa')$,
any integrable $\widetilde{p}(\chi)$ guarantees that the
expected degree of nodes with hidden variable $\kappa$ is indeed
$\kappa$, $\bar{k}(\kappa)=\kappa$, so that $\gamma$, which is a
model parameter, is indeed the exponent of the degree distribution
in generated networks.

We now want to map the expected degree $\kappa$ of each node to a
radial position $r$ within a disk of radius $R$, such that after the
mapping, the radial distribution of nodes is $\rho(r)\approx\alpha
e^{\alpha(r-R)}$, i.e., as in the hyperbolic $\mathbb{H}^2$ model
introduced above. To have this $\rho(r)$, we must select the $\kappa
\to r$ mapping according to
\begin{equation}\label{eq:kappa.vs.r}
\kappa = \kappa_0 e^{\frac{\zeta}{2}(R-r)},\;
\frac{\zeta}{2}=\frac{\alpha}{\gamma-1},\;
N=c\,e^{\frac{\zeta}{2}R},\;
c=\pi\mu\kappa_0^2,
\end{equation}
where $\zeta$ is fixed by the values of $\gamma$ and target
$\alpha$. We see that $\kappa(r)$ and consequently $\bar{k}(r)$
scale with $r$ as in the $\mathbb{H}^2$ model, while the connection
probability $\widetilde{p}(\chi)$ becomes
$\widetilde{p}\left(e^{\frac{\zeta}{2}(x-R)}\right)$, where the
effective distance
\begin{equation}\label{eq:x-effective}
x = r + r' + \frac{2}{\zeta}\ln\frac{\Delta\theta}{2}
\end{equation}
is approximately equal to the hyperbolic distance between the two
nodes in the disk. Indeed, the true hyperbolic distance $x$ between
two points with polar coordinates $(r,\theta)$ and $(r',\theta')$ in
the hyperbolic space $\mathbb{H}^2$ of curvature $K=-\zeta^2$ is
$\cosh\zeta x = \cosh\zeta r \cosh\zeta r' - \sinh\zeta r \sinh\zeta
r' \cos\Delta\theta$, which for sufficiently large $\zeta r$, $\zeta
r'$, and $\Delta\theta>2\sqrt{e^{-2\zeta r}+e^{-2\zeta r'}}$ is closely approximated by
\begin{equation}\label{eq:x-true}
x = r + r' + \frac{2}{\zeta}\ln\sin\frac{\Delta\theta}{2}.
\end{equation}
Since the effective and true hyperbolic distances in
Eqs.~(\ref{eq:x-effective},\ref{eq:x-true}) are approximately equal,
the value of $\zeta$ in Eq.~(\ref{eq:kappa.vs.r}) is indeed the
square root of curvature of the hyperbolic disc, in agreement with
Eq.~(\ref{eq:true-gamma}) in the $\mathbb{H}^2$ model. We also
notice that since the connection probability $\widetilde{p}(\chi)$
in the $\mathbb{S}^1$ model can be any integrable function, the
connection probability $p(x)$ in the $\mathbb{H}^2$ model can be
\emph{any function} of the form
$p(x)=\widetilde{p}\left(e^{\frac{\zeta}{2}(x-R)}\right)$.

Given this freedom of choice of the connection probability, let us
consider the family of functions
\begin{equation}\label{eq:fermi_function}
p(x) = \frac{1}{1+e^{\frac{\zeta}{2T}\left(x-R\right)}}
\end{equation}
parameterized by $T>0$. One motivation to focus on this family is
that it generates exponential random graphs in the statistical
mechanics sense~\cite{PaNe04}. Eq.~(\ref{eq:fermi_function}) is
nothing but the grand canonical Fermi-Dirac distribution, and $T$ is the
system temperature. From the physical perspective, graph edges are
non-interacting fermions with energies equal to their hidden
hyperbolic lengths, and $R$ is the chemical potential defined by the
condition that $\bar{k}N/2$, the number of edges-fermions, is fixed
on average. At $T\to0$ Eq.~(\ref{eq:fermi_function}) converges to
$p(x)=\Theta(R-x)$, which {\it a posteriori} justifies our choice of
the step function connection probability in the $\mathbb{H}^2$
model.

\begin{figure}
    \includegraphics[width=3.4in]{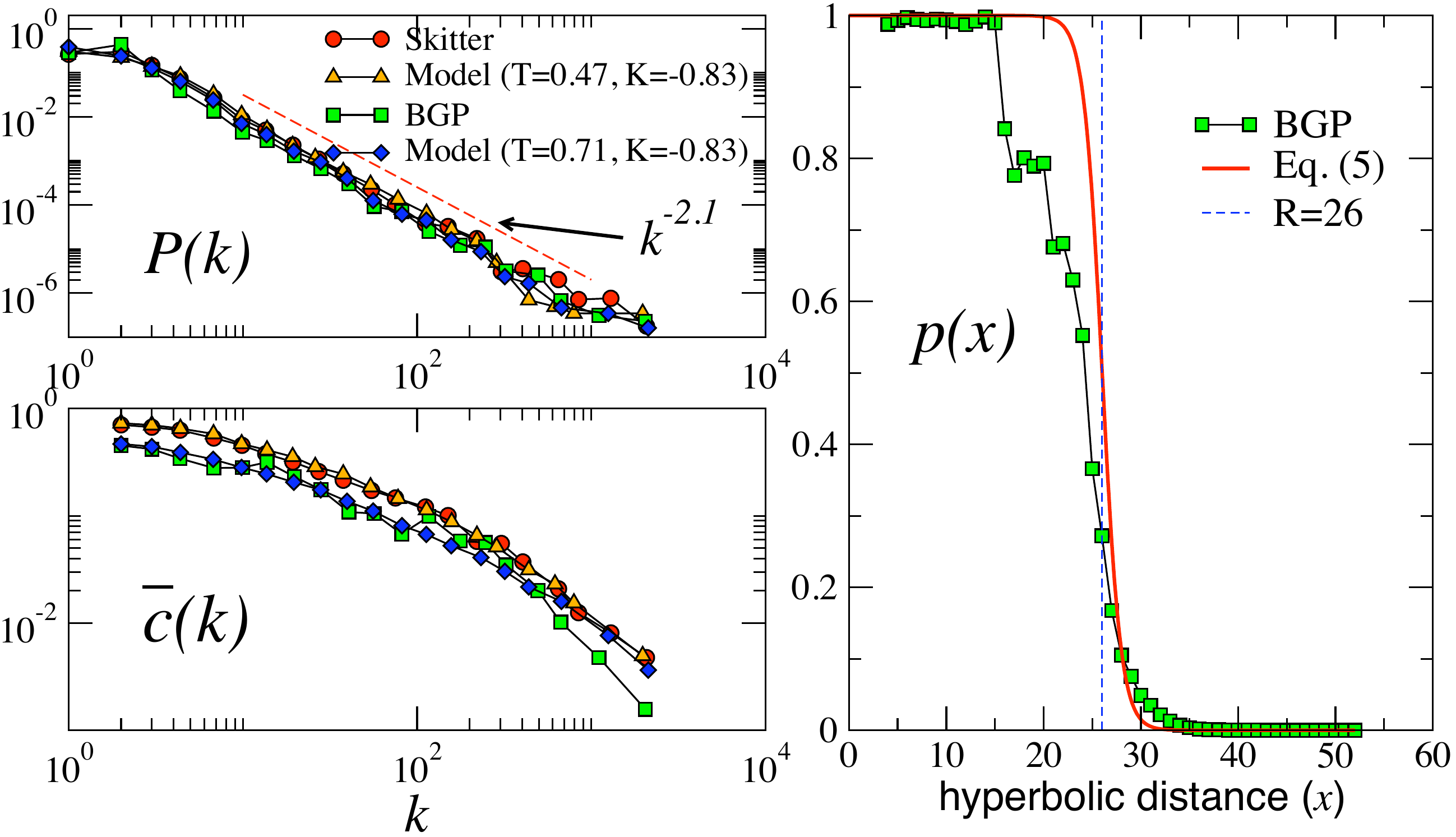}
    \caption{Networks in the $\mathbb{H}^2$ model vs.\ the Internet. {\bf Left:}
    The degree distribution $P(k)$ and degree-dependent clustering coefficient $\bar{c}(k)$ are shown for the skitter
    ($\bar{k}=6.29$, $\bar{C}=0.46$)
    and Border Gateway Protocol (BGP) ($\bar{k}=4.68$, $\bar{C}=0.29$)
    views of the Internet from~\cite{MaKrFo06},
    and for modeled networks with curvature $K=-0.83$ and two values of temperature $T$,
    $0.47$ ($\bar{k}=6.03$, $\bar{C}=0.44$) and $0.71$ ($\bar{k}=4.85$, $\bar{C}=0.25$).
    {\bf Right:} The empirical connection probability in the hyperbolically embedded Internet,
    compared to Eq.(\ref{eq:fermi_function}).
    \label{fig:H2_reality}
    }
\end{figure}

The dependence on temperature in the model is peculiar. At zero
temperature, the network is in the strongly degenerate ground state.
As we heat it up, particles explore higher-energy states, i.e.,
edges connect longer distances, which affects clustering. At
$T\to0$, clustering is maximized. It monotonically decreases with
$T$, and at $T\to1$ we have a phase transition with clustering going
to zero, and the network losing its cold-state metric structure. In
the cold regime with $T<1$, the exponent of the degree distribution
$\gamma$ depends only on the ratio $\alpha/\zeta$ via
Eq.~(\ref{eq:true-gamma}). Therefore, we can set $\alpha=1/2$
without loss of generality, so that $\gamma=1/\zeta+1$ is fully
defined by curvature $K>-1$. In the hot regime with $T>1$,
clustering remains zero, the chemical potential is no longer given
by $N=c\,e^{\frac{\zeta}{2}R}$ but by $N=c\,e^{\frac{\zeta}{2T}R}$,
and $\gamma$ also depends on temperature, $\gamma=T/\zeta+1$.
Therefore at $T\to\infty$ the graph ensemble is identical to
classical random graphs, as all fermions are uniformly distributed
across all energies, i.e., all pairs of nodes are connected with the
same probability independent of the hidden distance between them,
and the network loses its cold-state hierarchical structure.
Combining the cold and hot regimes,
\begin{equation}
\gamma =
\begin{cases}
1/\zeta+1 & \text{if $T<1$ and $\zeta<1$,}\\
T/\zeta+1 & \text{if $T>1$ and $\zeta<T$,}\\
2 & \text{otherwise}.
\end{cases}
\end{equation}
Finally, constant $c$ fixing the average degree in the network is
\begin{equation}
c \approx
\begin{cases}
\bar{k}\frac{\sin\pi T}{2T} \left(1-\zeta\right)^2 \approx
\kappa_0^2\frac{\sin\pi T}{2\bar{k}T}
& \text{if $T<1$,}\\
\bar{k}\left(\frac{\pi}{2}\right)^{\frac{1}{T}}\frac{T-1}{T^3}\left(T-\zeta\right)^2
\xrightarrow[T\to\infty]{}\bar{k}
& \text{if $T>1$.}
\end{cases}
\end{equation}
The $\mathbb{H}^2$ model can thus generate classical random graphs
and scale-free networks with any average degree, power-law exponent
$\gamma>2$, and clustering. In Fig.~\ref{fig:H2_reality}, left, we
see that the curvature and temperature of the Internet are
approximately $K=-0.83$ and $T=0.6\pm0.1$.

Eq.~(\ref{eq:kappa.vs.r}) establishes a formal equivalence between
the $\mathbb{S}^1$ and $\mathbb{H}^2$ models we introduced
in~\cite{SeKrBo08} and here. The two models generate similar network
topologies thanks to the similarity between the effective and true
hyperbolic distances in Eqs.~(\ref{eq:x-effective},\ref{eq:x-true}).
However, if we are to study other geometric properties of these
networks, such as their navigability~\cite{BoKrKc08}, then it does
matter a lot what distances, spherical $d$ native to $\mathbb{S}^1$
or hyperbolic $x$ native to $\mathbb{H}^2$, we use to navigate a
network. The latter distances $x$ are dominated by $r+r'$, minus
some small $\theta$-dependent corrections. This effect can be
observed in Fig.~\ref{fig:escher}, right, where we show some
hyperbolic geodesics between nodes in a small modeled network. These
geodesics follow closely the radial directions between the nodes and
the origin, i.e., they follow the same pattern as the shortest paths
in the embedded network. Spherical distances $d$ are at the other extreme, as
their gradient lines lie in the orthogonal tangential directions.

To demonstrate how such differences in distance calculations affect
the efficiency of transport processes on networks, we embed the
real Internet topology from~\cite{dimes-ccr}
into $\mathbb{H}^2$ using maximum-likelihood
techniques. Specifically, we first assign to nodes random angular
coordinates, while their radial coordinates are fixed by
Eq~(\ref{eq:kappa.vs.r}). We then execute the Metropolis-Hastings
algorithm~\cite{Newman99-book} by moving random nodes to new
locations with the same radial coordinate but with a randomly chosen
new angular coordinate. We accept each move with probability
$\min(1,\mathcal{L}_a/\mathcal{L}_b)$, where $\mathcal{L}_b$ and
$\mathcal{L}_a$ are the likelihoods, before and after the move, that
the network is produced by our $\mathbb{H}^2$ model with parameters
matching the Internet in Fig.~\ref{fig:H2_reality}, left. Formally,
${\cal L} = \prod_{i<j}p(x_{ij})^{a_{ij}}[1-p(x_{ij})]^{1-a_{ij}}$,
where $\{a_{ij}\}$ is the Internet adjacency matrix, and $x_{ij}$ is
the hyperbolic distance between nodes $i$ and $j$.

After this process has converged, we perform greedy
routing as in~\cite{BoKrKc08} in the resulting embedding. We randomly
select a source, and try to find a path to a random destination by
selecting the next node on the path as the current node's neighbor
closest to the destination in $\mathbb{H}^2$. This process can be
unsuccessful, as it can get stuck at intermediate nodes that have no
neighbors closer to the destination than themselves. The percentage
of successful greedy paths and their hop-length averaged over $10^5$
random source-destination pairs are $94.5\%$ and $3.95$ (the average
length of shortest paths is $3.46$). For comparison, the same
numbers using the $\mathbb{S}^1$ distances are $75.9\%$ and $4.29$.
The reason for the exceptionally high ratio of successful paths in
the $\mathbb{H}^2$ case is that the shortest paths in the Internet
stay close to the hyperbolic geodesics, followed by greedy
navigation, between the corresponding source and destination. In
other words, the real Internet topology is remarkably congruent with
underlying hyperbolic geometry.

Even more striking in this regard is Fig.~\ref{fig:H2_reality},
right, where we show the empirical connection probability for the
links vs.\ their hyperbolic distances in the embedded Internet,
juxtaposed with the theoretical connection probability in our
$\mathbb{H}^2$ model. The similarity between the two provides
empirical evidence that our model reflects reality. If the real
Internet were not congruent with our hyperbolic model, then no
maximum-likelihood technique used for its embedding would be able to
make it such.

In summary we have shown that complex network topologies are
congruent with hyperbolic geometries. We can start with hyperbolic
geometry, and show that it naturally gives rise to scale-free
topology, or we can start with the latter, and show that hyperbolic
geometry is its effective geometry. In this geometric approach,
clustering and heterogeneous degree distributions appear as simple
consequences of the metric and negative-curvature properties of
hyperbolic spaces. In our hyperbolic model, the space curvature
controls the heterogeneity of the degree distribution, while
clustering is a function of temperature. Fermi-Dirac statistics
provides a physical interpretation of hidden distances as energies
of the corresponding links-fermions. This analogy may contribute to
applications of the standard tools of statistical mechanics to the
analysis of complex networks~\cite{PaNe04}, which can be informally
thought of as negatively curved containers of ultracold fermions.
The Internet embedding, besides providing empirical evidence that
our model is realistic, shows that the efficiency of transport
processes without global knowledge is maximized if such processes
use (effective) hyperbolic distances. If networks evolve to be
efficient with respect to their functions---and transport is one of
such functions,---then this finding further supports our hyperbolic
metric space approach.

The Internet embedding may also prove practically useful, since
routing in it is extremely efficient and requires only local
information about hyperbolic coordinates of node neighbors. Global
knowledge of the large-scale Internet topology is a major
scalability bottleneck in Internet routing
today~\cite{iab-raws-report-phys}. Another potential application of
our work is protein folding, where hidden spaces are protein
conformation energy profiles, and the protein folding process is
greedy routing toward the minimum-energy
conformation~\cite{RaGna07}. Yet another class of applications
involves cases where to have a right model for similarity distances
is a key, such as recommender systems used by companies such as Amazon
or Netflix. Their efficiency depends on how accurately the
similarities between consumers are estimated. Our hyperbolic
explanation of the structure of complex networks is by no means the
only possible mechanism capable of generating scale-free topologies
with strong clustering. Therefore, the question of special interest
is whether our explanation is (implicitly) equivalent to existing
models, among which preferential attachment~\cite{BarAlb99} appears
to be most popular?

\begin{acknowledgments}
We thank  A.~Goltsev, S.~Dorogovtsev, A.~Samukhin,
R.~Pastor-Satorras, A.~Baronchelli, M.~Newman, J.~Kleinberg,
Z.~Toroczkai, F.~Menczer, A.~Clauset, D.~Clark, K.~Fall, kc claffy,
and others for useful discussions and suggestions. This work was
supported by NSF CNS-0434996, CNS-0722070, DHS N66001-08-C-2029,
Cisco Systems, and FIS2007-66485-C02-02.
\end{acknowledgments}


\end{document}